
\font\ninerm=cmr9   \font\twerm=cmr12

\magnification 1200
\footline={\hfill}
\rightline{ SLAC-PUB-6688}
\rightline{ October 1994}
\rightline{ T/E}
\null\vskip .5truecm
\centerline{\twerm SEMI-EXCLUSIVE PION PRODUCTION}
\vskip 1mm
\centerline{\twerm IN DEEP-INELASTIC SCATTERING}
\vskip 2truecm
\centerline{\twerm A. Brandenburg\footnote{$^{1}$}{\ninerm \noindent
Max Kade
fellow. Address after November, 1st: Institut
f\"ur Theoretische Physik E, Physik\-zentrum, RWTH Aachen, D-52056
Aachen, Germany
}$^{,}$\footnote{$^{2}$}
{\ninerm \noindent Work supported by the
Department of Energy, contract DE-AC03-76SF00515},
{}~ V. V. Khoze$^{2,}$\footnote{$^{3}$}{\ninerm \noindent
Address
after November, 1st: University of Durham, Dept. of Physics,
Science Labs., Durham, DH1 3LE, England
} ~ and D. M\"uller$^{2,}
$\footnote{$^{4}$}{\ninerm \noindent
Supported
by Deutscher Akademischer Austauschdienst}}
\vskip .5truecm
\centerline{\it Stanford Linear Accelerator Center}
\centerline{\it Stanford University, Stanford, California 94309}
\vskip 2truecm
\centerline{\twerm Abstract}
{\noindent We calculate azimuthal asymmetries and the
Callan-Gross $R$-ratio
for semi-exclusive pion production in deep inelastic scattering
taking into account higher twist effects.
Our results are qualitatively different from the QCD-improved parton
model predictions for semi-inclusive deep inelastic
scattering.}

\vskip 3truecm
\centerline{\it Submitted to Phys. Lett. {\rm B}}

\vfil\eject
\pageno=1
\footline={\hss\tenrm\folio\hss}
\baselineskip=6mm

1. The semi-inclusive deep inelastic process
$$ \ell + p\to \ell ' +h+X ,\eqno(1)
$$
\noindent where $\ell$ and $\ell '$ are charged
leptons and $h$ is a detected
hadron, has been recognized long ago [1] as an
important testing ground for QCD.
In particular, measurements of the azimuthal angle
$\varphi$ of the observed hadron
provide additional information on the underlying QCD
production
mechanism. The angle $\varphi$ is defined in the target rest frame
as follows:
The $z$ axis is chosen in the direction opposite to the
three-momentum transfer
${\bf q}$ of the leptons, and the $x$-$z$ plane is the lepton
scattering plane
with the lepton three momenta having positive $x$ components.
The angle $\varphi$
is the azimuthal angle of the observed hadron  about the $z$ axis.

\par
Different mechanisms to generate
azimuthal asymmetries -- $\langle \cos\varphi\rangle$ and $\langle
\cos 2\varphi
\rangle$ -- have been discussed in the
literature. Georgi and Politzer [1] found
a negative contribution to $\langle \cos\varphi\rangle$ in the
QCD-improved parton
model and proposed the measurement of this quantity as a clean test of
perturbative QCD.
Cahn [2] took into account the effects due to the
intrinsic transverse momenta
of partons bound inside the proton.
This nonperturbative effect produces a negative
contribution to $\langle \cos\varphi \rangle$ and  a positive one to
$\langle \cos 2\varphi \rangle$.
More recently Chay, Ellis and Stirling
[3] combined perturbative and nonperturbative mechanisms
and were able to fit
the measured values for $\langle \cos\varphi \rangle$ [4].
The nonperturbative
effects were parameterized in Ref. [3]
by Gaussian distributions for the intrinsic momenta
of both the target (proton) and the observed hadron (pion).
In fact, the
hadronization effects cannot be neglected for the fit
of Ref. [3] to the data. The
importance of hadronization effects for the azimuthal
asymmetry was stressed by K\"onig and Kroll [5] and also by Berger [6].
The calculation
of Berger is made specifically for the case of single pion production
and takes into account
pion bound-state effects.

\par
It is an intriguing
feature of Berger's higher twist
mechanism that it generates azimuthal asymmetries
opposite in sign to
those of [3,2,1,5]. However, the competing mechanisms make it difficult
to give reliable
theoretical predictions for the azimuthal asymmetries and their
dependence on the whole set of kinematic variables of the process (1).
In this paper we reconsider
Berger's mechanism and discuss ways of isolating its
effects from those of the competing mechanisms discussed above.
Our motivation is twofold:
first, this approach calculates and incorporates in a nontrivial
way higher twist effects by relating them to the distribution amplitude
of the produced hadron; second, a similar approach for
Drell-Yan process was shown in [7] to be of importance in resolving
a long standing problem in
explaining the data for the angular distribution of the produced lepton.

\par
2. First we discuss the kinematics of process (1). Let $l$ and $l'$
be the four-momenta of the leptons
with $q=l-l'$ being the four-momentum transfer and
$P_N$ and $P_h$
those of the proton and the observed hadron. We define the usual
set of kinematic variables
$$ x_B={Q^2\over 2P_N\cdot q},\ \ \ \ \  y={P_N\cdot
q\over P_N\cdot l},\ \ \ \ \ z={P_h\cdot P_N\over q\cdot P_N}
,\eqno (2)$$
where $Q^2=-q^2$.
We will also use $Q=\sqrt{Q^2}$
and $\nu=q^0,\  E=l^0$ defined
in the proton rest frame.
The  transverse momentum
of the observed hadron is given by ${\bf P}_h^T = p_T \
(\cos \varphi, \sin \varphi)$,
where $\varphi$ is the azimuthal angle introduced in the
previous section.
The differential cross section for semi-inclusive deep
inelastic scattering
is given by
$${2 P_h^0 \ d \sigma \over d^3 P_h dx_B dy} \
= \ {2 \pi \alpha^2 M y \over
Q^4} \ L_{\mu \nu} W^{\mu \nu},\eqno(3)$$
where $M$ is the proton mass and $ L_{\mu \nu}$ and
$ W^{\mu \nu}$ are the leptonic and hadronic tensors respectively.
It is convenient
to introduce helicity structure functions,
$$ W_L = \epsilon_L^{\mu} W_{\mu \nu}
\epsilon_L^{\nu}, $$
$$ W_T = {1 \over 2}\bigl(   \epsilon_x^{\mu} W_{\mu \nu}
\epsilon_x^{\nu}
+ \epsilon_y^{\mu} W_{\mu \nu} \epsilon_y^{\nu} \bigr), $$
$$ W_{LT} = -\bigl(   \epsilon_x^{\mu} W_{\mu \nu} \epsilon_L^{\nu}
+ \epsilon_L^{\mu}
W_{\mu \nu} \epsilon_x^{\nu} \bigr), $$
$$ W_{TT} = {1 \over 2}\bigl(
\epsilon_x^{\mu} W_{\mu \nu} \epsilon_x^{\nu}  -
\epsilon_y^{\mu} W_{\mu \nu} \epsilon_y^{\nu} \bigr), \eqno(4) $$
where $\epsilon_x$, $\epsilon_y$ and $\epsilon_L$
are polarization vectors
for the vector boson polarized respectively in the
$x$, $y$ and $z$ directions.
Following Ref. [8] we introduce the following hadronic
structure functions,
$$ H_1 \ = \ {M\over 2 z} \ W_T,$$
$$ H_2 \ = \ {\nu Q^2 \over 2 z |{\bf q}|^2} \ (W_L + W_T),$$
$$ H_3 \ = \ { Q^3 \over 2 z |
{\bf q}|p_T} \ W_{LT},$$
$$ H_4 \ = \ { Q^4 \over 2 z p_T^2 \nu} \ W_{TT}.\eqno(5)$$
The differential cross section is conveniently
expressed in terms of these structure functions [8], namely in the deep
inelastic
limit we get
$$ { Q^2 \ d \sigma \over d x_B dy d z d p_T^2 d \varphi} \ =
\ { 4 \pi \alpha^2 ME \over Q^2}
\ \Bigl( x_B y^2 \ H_1 \ + \ (1-y) \ H_2 $$
$$ \ + \ {p_T \over Q}(2-y) \sqrt{1-y} \ H_3 \ \cos \varphi \ + \
{p_T^2 \over Q^2} (1-y)
\ H_4 \ \cos 2\varphi \ \Bigr), \eqno(6)$$
where $ H_i = H_i(x_B, z, p_T^2, Q^2)$.\par
This decomposition exhauts the kinematical information for reaction
(1).
Apart from the overall normalization factor one may specify
three independent observables. Two of them we take to be the
angular asymmetries,
$$ \langle \cos \varphi \rangle \ = \ {1 \over 2} \
{p_T \over Q} \    {(2-y) \sqrt{1-y}
\ H_3 \over x_B y^2 \ H_1 \ + \ (1-y) \ H_2}
,\eqno(7)$$
$$ \langle \cos 2\varphi \rangle \
= \ {1 \over 2} \ {p_T^2 \over Q^2}    \
{(1-y) \ H_4 \over x_B y^2 \ H_1 \ + \ (1-y) \ H_2}.\eqno(8)$$
For the third observable we propose
$$ R \ = \ {H_2 \ - \ 2x_B \ H_1 \over H_2}
,\eqno(9)$$
which is the analog of the Callan-Gross $R$-ratio [9] in inclusive
deep inelastic scattering.

\par
3. We will now calculate observables (7-9)
for the specific case of
{\it semi-exclusive} pion production. More precisely, we want to
consider a process in which a
direct pion with $z$ close to $1$ observed
in charged lepton -- proton scattering is not surrounded by
other high-$p_T$ hadrons. In other words, the pion
is not a member of a jet.
The events we are interested in thus form only a small part
of the whole semi-inclusive
sample.
If none of the specifications above apply to the observed hadron
it is expected that the usual QCD-improved parton model diagrams
of Refs. [1,3] would give the dominant contribution.
Most of the experiments performed [4,10] did not make these
specifications\footnote{$^*$}{\ninerm Note that in the first paper of
Ref. [10] single pions were identified and a nonzero asymmetry was
observed. However, the data were not fitted to the general form of the
$\varphi$-dependence given in  eq. (6). Thus it is difficult
to draw conclusions
about the asymmetries defined in eqs. (7), (8).}.

\par
The dominant contribution to the {\it semi-exclusive} {\it pion}
production described above is given by the diagrams of Fig. 1 a,b.
Here the amplitude ${\cal A}$ for the reaction
$$ u+e^{-} \rightarrow e^{-} + \pi^{-} + d,\eqno(10)$$
is obtained [11,6,7] by convoluting  the hard
scattering partonic amplitude
$T(u+e^{-} \rightarrow e^{-} + u\overline{d} + d)$ with the pion {\it
distribution amplitude} $
\phi(\xi)$,
$$ {\cal A} \ = \ \int_0^1 d\xi \ \phi(\xi) \ T .\eqno(11)$$
This factorized form of the amplitude should be valid
for $z$ close to $1$.\par
For the hadronic cross section (6) we have to convolute the
cross section obtained
from the amplitude of eq. (11) with the parton distribution function
of the proton
(see [6,7] for more details). A typical contribution to the
hadronic tensor
$W_{\mu \nu}$, that is to the hadronic structure functions
$H_i$, is depicted in Fig. 1 c.

We note that no {\it intrinsic} transverse momentum for the {\it pion}
constituents is introduced in our model. The source of the observed
$p_T$ of the pion is the gluon exchange, see Fig 1.
For the sake of simplicity of our presentation the intrinsic transverse
momentum of the {\it proton} constituents will be neglected
in the formulae below, but will be discussed later.

\par
For the structure functions of eq. (5) we find
$$ H_1 \ = \ {\cal N} \ {1\over 2x_B} \
\Bigl( \ [I_2(z,p_T/Q)]^2 \ + \
{p_T^4 \over Q^4} [I_1(z,p_T/Q)]^2 \ \Bigr) ,$$
$$ H_2 \ = \ {\cal N}  \ \Bigl( \ [I_2(z,p_T/Q)]^2 \
+ \   4{p_T^2 \over Q^2}z^2
[I_1(z,p_T/Q)]^2 \ + \   {p_T^4 \over Q^4}
[I_1(z,p_T/Q)]^2 \ \Bigr) ,$$
$$ H_3 \ = \ {\cal N} \ 2z \ I_1(z,p_T/Q) \
\Bigl( \ I_2(z,p_T/Q) \ - \
{p_T^2 \over Q^2} I_1(z,p_T/Q) \ \Bigr) ,$$
$$ H_4 \ = \ -{\cal N} \ 2 \ I_1(z,p_T/Q) \ I_2(z,p_T/Q)
,\eqno(12)$$
where  $I_1(z,p_T/Q)$
and $I_2(z,p_T/Q)$
are defined as follows,
$$ I_1(z,p_T/Q) \ = \ z \ \int_0^1 d\xi
{\phi(\xi) \over {z \ - \ \xi(z^2-p_T^2/Q^2)}}   , $$
$$I_2(z,p_T/Q) \ = \ \int_0^1 d\xi {\phi(\xi) \over 1-\xi} \
- \ z^2 I_1(z,p_T/Q).
\eqno(13)$$
Since we will only be interested in the ratios (7)--(9), the overall
factor ${\cal N}$ is irrelevant.
\par Using crossing invariance
the results for $H_i$ can also be
obtained by analytic continuation
from the formulae
for the angular distribution coefficients
for the Drell-Yan
process given in [7].
The transformation formulae are given in the Appendix.
\par
Substituting the expressions for $H_i$ given in eq. (12) in eqs. (7)--(9)
we
obtain predictions for our observables
for a specified distribution amplitude $\phi(\xi)$. Since the expression
for
$H_3$ turns out to be positive and the one for $H_4$ to be
 negative, we find $\langle \cos
\varphi \rangle > 0 $ and  $\langle \cos 2\varphi \rangle < 0 $. \par
Now we want to discuss the effects of intrinsic transverse
momenta $\bf k_T$
of partons bound inside the proton, since from the work of
Cahn [2] we know
that such effects alone produce asymmetries opposite in
sign to our results.
We introduce intrinsic ${\bf k}_T$ by modifying the
distribution function of
the proton, $$ G_{q/N}(x,Q^2)\to d^2k_T {1 \over 4\ \langle k_T
\rangle^2}
\
\exp \left\{ -{\pi\ {\bf k}_T^2\over 4\ \langle k_T \rangle^2}\right\}
 G_{q/N}(x,Q^2).\eqno (14) $$
 In the diagrams of Fig. 1 the transverse momentum ${\bf p}_q^T$
 of the quark from the proton is now set equal to ${\bf k}_T$
 and the quark
 is off-shell. More precisely we use the parameterization
  $p_q^+=x\ P_N^+,\ p_q^-=0,\ {\bf p}_q^T={\bf k}_{T}$.
  The use of different prescriptions will modify our results in
  subleading
  orders of $k_T/Q$. In the calculation we use the mean value theorem for
  the integration over the absolute value of ${\bf k}_T$ and neglect
  subleading terms in $k_T/Q$.\par
  We plot our results as a function of $z$
  for three different values of $\langle {k}_{T}
  \rangle $.
  In Fig. 2 we plot  the observables  of eqs. (7)--(9) for $Q=2.5$ GeV,
  $p_T= 0.5$ GeV and $y=0.4$. For the distribution amplitude
  of the pion we choose a
  two-humped form, $\phi(\xi)=30(1-\xi)\xi(1-4(1-\xi)\xi)$ [12].
  Here we neglect the evolution of $\phi(\xi)$ with $Q$.
   The solid line corresponds to
  $\langle {k}_{T} \rangle=0$, the dashed line to  $\langle {k}_{T}
  \rangle=0.25$ GeV and the dash-dotted line to
  $\langle {k}_{T} \rangle=0.5$
  GeV. In Fig. 3 we plot the same observables for $Q=5$ GeV
  and $p_T=1$ GeV.
   The  choices of  $\langle {k}_{T} \rangle $ and $y$  are the
   same as in Fig. 2.
    Figs. 2 and 3 depict our main results.

    \par
    An important feature of our model is a steep increase of the
    $R$-ratio as $z\to 1$. This is analogous to the steep decrease
    of the $\lambda$ parameter in the end point region
    in the Drell-Yan process predicted
    by the same higher twist model
      [6,7]. This behavior of $\lambda$ was
     observed experimentally [13].

    \par
    As expected, a nonzero  $\langle {k}_{T} \rangle $ causes a change in
  the prediction of our model for the azimuthal asymmetries and the
  Callan-Gross $R$-ratio. This change becomes insignificiant
  for higher values
  of $Q$, see Fig. 3.
  For $Q$ high enough, the analytic expressions (12), (7)--(9) can be used
  even for nonzero  $\langle {k}_{T} \rangle $.
  \par
  In Fig. 4 we show the dependence of the $R$-ratio and of the
  azimuthal asymmetry $\langle \cos\varphi \rangle$ on different
  distribution amplitudes assuming  vanishing $\langle {k}_{T} \rangle$,
  $p_T/Q=0.3$ and $y=0.4$. To demonstrate that our predictions depend
  essentially only on the value of $ \langle (1-\xi)^{-1} \rangle
  \equiv
    \int_0^1 d\xi \phi(\xi)
  (1-\xi)^{-1} $ and not on the specific
  shape of the distribution amplitude,
   we choose the two-humped distribution amplitude of
  Figs. 2, 3 (solid line in Fig. 4) and
  a convex amplitude
  $\phi(\xi)=[(1-\xi)\xi]^{1/3}[{\rm B} (4/3,4/3)]^{-1}$
  (dashed line)  which both have
   $ \langle (1-\xi)^{-1} \rangle=5$.
   Here ${\rm B}(x,y)$ denotes the Beta function.
   For comparison we plot results
   also for the asymptotic amplitude $\phi(\xi)=6(1-\xi)\xi$
   (dash-dotted line) with
    $ \langle (1-\xi)^{-1} \rangle=3$.
  The magnitude of the $R$-ratio in the endpoint region and the
  value
  of  $\langle \cos\varphi \rangle$ are decreasing with increasing
   $ \langle (1-\xi)^{-1} \rangle$ and
      are not sensitive to the shape of the distribution
   amplitude.

     \bigskip
  As demonstrated, higher twist effects may be isolated in semi-exclusive
  pion production for moderate values of $Q$.
  We therefore look forward to future precision experiments.

\bigskip
{\bf Acknowledgments}

\noindent We wish to thank S. J. Brodsky and W. J. Stirling
for discussions.
\vfil \eject
{\twerm{\bf Appendix}}\vskip 0.5cm\noindent
By using crossing invariance we can relate the hadronic structure
functions $H_i$ for semi-inclusive deep inelastic
scattering defined in eq. (6) with the angular coefficients for
the Drell-Yan process
defined in the Gottfried-Jackson frame by
$$ {q^2d\sigma \over dq^2dq_T^2dxd\Omega}
\sim(1+ \lambda \cos^2\theta
+\mu\sin 2\theta\cos\phi+{\nu\over 2} \sin^2\theta\cos 2\phi)
, \eqno(A.1)$$
where the angular distribution coefficients $\lambda$, $\mu$ and $\nu$
are  functions of the kinematic variables $x=(q\cdot P_N)/
(P_{\pi}\cdot P_N)$ and $q_T^2/q^2$
. Here, $P_{\pi}$ and $P_N$ are the momenta of the initial
state hadrons and $q$ is the four-vector of the produced photon with
transverse momentum $q_T$.

\par\noindent
We get the following expressions relating
coefficients $\lambda$, $\mu$ and $\nu$ of the Drell-Yan process
with the structure functions $H_i$ of eq. (5):
$$x_B\ H_1(z,p_T/Q)=-N\ [1+\lambda(x',\rho)-\Delta(x',\rho)]/2
$$
$$ H_2(z,p_T/Q)=-2\ N\ [\lambda(x',\rho)-{3\over
2}\Delta(x',\rho)]$$
$$ p_T/Q\ H_3(z,p_T/Q)=\sqrt{-\rho^2}\ N\ [-2{\mu(x',\rho)\over
\sqrt{\rho^2}}+\Delta '(x',\rho)]$$
$$p_T^2/Q^2\ H_4(z,p_T/Q)=N\ [\nu(x',\rho)+\Delta(x',\rho)] \eqno(A.2)$$
with
$$ \Delta(x',\rho)={4\rho^2\over (1+\rho^2)^2}
[\lambda(x',\rho)-\nu(x',\rho)/2-{(1-\rho^2)\over
\sqrt{\rho^2}}\mu(x',\rho)],$$
$$\Delta '(x',\rho)={4\over (1+\rho^2)^2}
[(1-\rho^2)(\lambda(x',\rho)-\nu(x',\rho)/2)+4
\sqrt{\rho^2}\mu(x',\rho)] \eqno(A.3)$$
and
$$x'(z)={1\over z}\ ,\ \ \ \ \ \rho(z,p_T/Q)={i\
p_T/Q\over z}\ ,\ \ \ \ \
  \eqno(A.4)$$
 and $N$ is an overall normalization factor.

\vfil\eject

{\twerm{\bf  References}}
\vskip 0.5cm
\item{[1]} H. Georgi and H.D. Politzer, {\it Phys. Rev. Lett. }
{\bf 40}(1978) 3.
\item{[2]} R.N. Cahn, {\it Phys.Lett. }{\bf 78B}; {\it Phys. Rev. }
{\bf D 40}(1989) 3107.
\item{[3]} J. Chay, S.D. Ellis and W.J. Stirling, {\it Phys. Rev. }
{\bf D 45}(1992) 46.
\item{[4]} M.R. Adams et al., {\it Phys. Rev. }{\bf D 48} (1993) 5057;
D.M. Jansen, Ph. D. thesis, University of Washington, 1991.
\item{[5]} A. K\"onig and P. Kroll, {\it Z. Phys. }{\bf C 16} (1982) 89.
\item{[6]} E.L. Berger, {\it Z. Phys. }{\bf C 4} (1980) 289.
\item{[7]} A. Brandenburg, S.J. Brodsky, V.V. Khoze and
D. M\"uller, {\it Phys. Rev. Lett. }{\bf 73} (1994) 939.
\item{[8]} P.J. Mulders, {\it Phys. Rep. }{\bf 185}
(1990) 83; J. Levelt and P.J. Mulders, {\it Phys. Rev. }
{\bf D 49} (1994) 96.
\item{[9]} C.G. Callan and D.J. Gross, {\it Phys. Rev. Lett. }
{\bf 22} (1969) 156.
\item{[10]} K.C. Moffeit et al., {\it Phys. Rev. }
{\bf D 5} (1972) 1603;
\item{} V. Eckart et al., {\it Nucl. Phys. }{\bf B 55} (1973) 45;
\item{} J.T. Dakin et al., {\it Phys. Rev. Lett. }
{\bf 29} (1972) 746;
\item{} B. Gibbard et al., {\it Phys. Rev. }
{\bf D 11} (1975) 2367;
\item{} C. Tao et al., {\it Phys. Rev. Lett. } {\bf 44} (1980) 1726;
\item{} M. Arneodo et al., {\it Z. Phys. }{\bf C 34} (1987) 277;
\item{} see also {\it High $p_T$ physics and higher twists},
eds. M. Benayoun, M. Fontannaz and J.L. Narjoux, {\it Nucl.
Phys. B (Proc. Suppl.) }
{\bf 7 B} (1989).
\item {[11]} G.P. Lepage and S.J. Brodsky, {\it Phys. Rev. }{\bf D 22}
(1980) 2157.
\item{[12]} V.L. Chernyak and A.R. Zhitnitsky, {\it Phys. Rep. }
{\bf 112}(1984) 173.
\item{[13]} J.G. Heinrich et al., {\it Phys. Rev.} {\bf D 44}(1991) 44
{}.
\vfil\eject
{\twerm{\bf  Figure Captions}}
\vskip 0.5cm

  \item{\rm Fig. 1:}Diagrams (a) and (b) give the leading contribution
  to the amplitude of reaction (10). Diagram (c) gives a typical (one
  out of four) contribution to the hadronic tensor $W^{\mu\nu}$.

  \bigskip

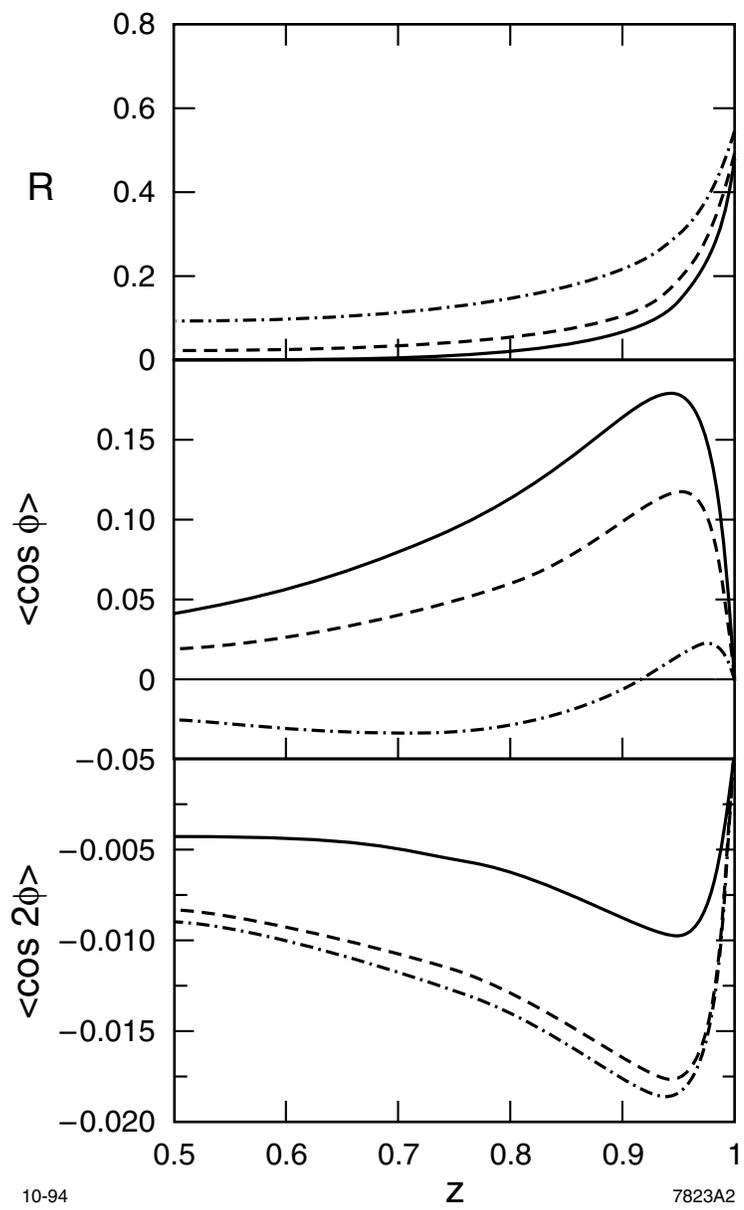
  \item{\rm Fig. 2:}The observables $R$, $\langle \cos\varphi \rangle$
  and $\langle \cos 2\varphi \rangle$ defined in eqs. (7)--(9) vs. $z$
  for $Q=2.5$ GeV, $p_T=0.5$ GeV, $y=0.4$ and
  $\phi(\xi)=30(1-\xi)\xi(1-4(1-\xi)\xi)$. The solid line corresponds
  to $\langle k_T \rangle=0$, the dashed line to $\langle k_T
  \rangle=0.25$ GeV and the dash-dotted line to $\langle k_T \rangle=0.5$
  GeV.
 \bigskip

  \item{\rm Fig. 3:}The same quantities as in Fig.2 for $Q=5$ GeV and
  $p_T=1$ GeV. The choices for $\langle k_T \rangle$  and $y$
  are the same as in
  Fig. 2.
\bigskip

  \item{\rm Fig. 4:}The $R$-ratio and $\langle \cos\varphi \rangle$ as
  a function of $z$ for $\langle k_T \rangle=0$, $p_T/Q=0.3$ and
  $y=0.4$. The results are plotted
  for three different distribution amplitudes. The solid line corresponds to
  the two-humped distribution amplitude also used in Figs. 2, 3. The
  dashed line corresponds to the convex amplitude
  $\phi(\xi)=[(1-\xi)\xi]^{1/3}[B(4/3,4/3)]^{-1}$ and the dash-dotted
  line to the asymptotic one $\phi(\xi)=6(1-\xi)\xi$.
\end